\documentstyle[12pt]{article}

\input epsf

\newcommand{\sub}[1]{_{\mbox{\scriptsize#1}}}

\newcommand\ee{\end{equation}}
\newcommand\be{\begin{equation}}
\newcommand\eea{\end{eqnarray}}
\newcommand\bea{\begin{eqnarray}}

\newcommand\TeV{\,\mbox{TeV}}
\newcommand\GeV{\,\mbox{GeV}}
\newcommand\MeV{\,\mbox{MeV}}

\newcommand\mpl{m_{Pl}}

\newcommand\lsim{\mathrel{\rlap{\lower4pt\hbox{\hskip1pt$\sim$}}
    \raise1pt\hbox{$<$}}}
\newcommand\gsim{\mathrel{\rlap{\lower4pt\hbox{\hskip1pt$\sim$}}
    \raise1pt\hbox{$>$}}}

\def\m{\mbox m}

\def\pd{\partial}

\begin{document}
%\begin{titlepage}
\rightline{LANCASTER-TH 95/07, CLNS 95/1384, SUSX-TH 96/31, hep-ph/9602263}
\vspace*{3mm}
\begin{center}
{\large\bf Some aspects of thermal inflation:\break
the finite temperature potential and topological defects}
\end{center}
\bigskip
\begin{center}
{\large\rm Tiago Barreiro$^1$, 
E. J.~Copeland$^1$, David H.~Lyth$^2$}
\end{center}
\begin{center}
       {\large\rm and}
\end{center}
\centerline{\large\rm Tomislav Prokopec$^3$}
\medskip
%\address
%\begin{center}
\centerline{$^1$School of Mathematical and Physical Sciences,}
\centerline{\it	University of Sussex, Falmer, Brighton BN1 9QH,~~U.~K.}
%\end{center}
%\address{\em $^1$School of Mathematical and Physical Sciences,
%University of Sussex, Falmer, Brighton BN1 9QH,~~U.~K.}
%\address
\centerline{\it $^2$School of Physics and Chemistry, Lancaster University,
Lancaster LA1 4YB,~~U.~K.}
%\address
\centerline{\it $^3$Newman Lab for Nuclear Studies, Cornell University,
Ithaca NY14853, USA}
\date{\today}
%\maketitle

\bigskip

\begin{center}
{\bf Abstract}
\end{center}

\baselineskip 16pt

Currently favoured extensions of the Standard Model typically
contain  `flaton fields' defined as fields with large vacuum 
expectation 
values (vevs) and almost flat potentials.
If a flaton field is trapped at the origin in the early
universe, one expects `thermal inflation' to
take place before it rolls away to the true vacuum,
because the finite-temperature correction to the potential
will hold it at the origin until the temperature falls below
$1\TeV$ or so. In the first part of the paper, that expectation
is confirmed by an estimate of the finite temperature corrections
and of the tunneling rate to the true vacuum, paying careful attention
to the validity of the approximations that are used. The second part 
of the paper considers topological defects which may be  produced at 
the end of an era of thermal inflation. If the flaton fields 
associated with the era are GUT higgs fields, then its end corresponds
to the GUT phase transition. In that case, the abundance of monopoles and 
of GUT higgs particles will have to be diluted by a second era of 
thermal
inflation (or perhaps some non-thermal analogue).
 Such an era will not affect the cosmology of GUT strings,
for which the crucial parameter is the string mass per unit length.
Because of the flat Higgs potential, the
GUT symmetry breaking scale required for the strings to be
a candidate for the origin of large scale structure and the cmb 
anisotropy is  about three times bigger than usual, but given the 
uncertainties it is still compatible with the one 
required by the unification of the Standard Model gauge
couplings.
The cosmology of textures and of global 
monopoles is unaffected by the flatness of the potential.

\bigskip

\begin{flushleft}
\end{flushleft}

%\end{titlepage}

\vfil
\eject

\baselineskip 16pt

\section{Introduction}

In presently favoured extensions of the Standard Model,
the space of the scalar fields contains many
directions in which the potential is almost flat out to large field
values.\footnote{Flat directions occur rather generically 
in supersymmetric theories as a consequence of a non-renormalization
theorem \cite{GrisaruSiegelRocek}, \cite{Seiberg}. 
The renormalizability requirement implies that there are flat 
directions protected to all orders in perturbation theory. 
The terms allowed then have couplings suppressed by some power 
of the Planck mass.}
In some of these directions the mass-squared is likely to be
negative, leading to a nonzero vev whose magnitude will be large 
because
the potential is flat. Fields of this kind, with  large vevs
and almost flat potentials, have been called 
flatons \cite{decay},
and it has long been recognized that they may be cosmologically
significant
\cite{problem,dinefisch,coughlan,decay,yam,therm,Enqvist,Ross,yam2,interm}.
\footnote
{Note the etymology. The term `flaton' refers to the {\em flat\/}
potential,
not to in{\em flat\/}ion. Conversely, the familiar word `inflaton'
refers to
the field which is slowly rolling during ordinary
{\em inflat\/}ion. The term `flaton' was invented long before 
it was realized that such a field can give rise to a different type 
of 
inflation (thermal inflation).}

Recently a previously almost \cite{therm} unnoticed aspect of
flaton cosmology was studied, and termed thermal inflation
\cite{gutti,therminf}. 
Thermal inflation 
is made possible by the flatness of the potential, which
near the origin allows the (positive) finite temperature 
contribution to the mass-squared 
to dominate the true (negative) mass-squared long after the 
thermal contribution to the energy density has become negligible.
If it occurs, thermal inflation 
completely alters the standard cosmology, and one purpose of this 
paper
is to examine carefully the features of the finite temperature
potential that are supposed to lead to it. 
Taking into account the validity of the various approximations,
we shall verify that thermal inflation indeed takes place,
provided that the field is trapped at the origin
and that it has sufficiently strong interactions with
light fields.

When thermal inflation ends, and the field rolls away from the origin,
topological defects may be produced
just as in the more familiar case of a non-flat potential.
The second purpose
of this paper is to see what effect the flatness of the potential has
on these defects, and their cosmology. We are particularly interested 
in 
the case that the GUT higgs particles are flatons,
so that the defects produced are monopoles and (depending on the GUT)
cosmic strings, but we shall also study the case of global symmetry
breaking.

Recently it has been pointed out \cite{klslatest} that the 
interaction of other scalar fields with the flaton
might cause inflation even without thermalization.
In that case the situation is more model dependent and we shall not 
study it in this paper, though much of our discussion of topological 
defects 
will still be applicable.

Thermal inflation is expected
to take place after ordinary (slow-roll) inflation, and to last for
fewer $e$-folds. 
Such a late epoch of inflation can play a welcome role
in diluting the abundance of unwanted relics, while leaving unaltered
the large-scale density perturbation accounting for the cmb anisotropy
and large scale structure. (As we shall discuss this remark applies
whether the perturbation is caused by a  vacuum fluctuation
during ordinary inflation, or by topological defects such as strings
or textures.) There can be more than one epoch of thermal inflation, 
so that a second epoch can dilute relics left over from
a first epoch, and we shall see that this is indeed essential
if one is to have a GUT phase transition with a flat potential.

Before getting into more detail we need to be 
precise about what is meant
by a `large' vev, and a potential which is `almost flat'.
In this paper we take these terms  to be defined 
with respect to the energy scale $10^2$ to
$10^3\,{\rm GeV}$, which is the scale of supersymmetry breaking as 
defined by
the masses of the supersymmetric partners of known particles 
\cite{susy}. Thus the vev $M$ of a flaton field is much bigger than
the above scale, whereas the the energy scale 
$|V''|^{1/2}$ defined by the curvature of its potential is 
taken to be of
order that scale. In particular the mass $m$ of a flaton particle 
is taken to be of that order. 
Since the potential as well as its first derivative 
vanishes at the vev, 
its value $V_0$ at the origin is of order $m^2M^2$.

Each flaton is either a gauge singlet
or a GUT Higgs field. 
A GUT is nowadays considered optional
because it is difficult to implement, and can perhaps be avoided by 
appealing to superstring unification near the Planck scale.
On the other hand the observed Standard Model couplings 
are compatible with the
existence of a GUT, with Higgs vevs of order $10^{16}\GeV$
(plus possible additional intermediate scale vevs \cite{Mohapatra}).
If one accepts the existence of a GUT then it is attractive to 
suppose that
the Higgs fields are flatons because one might then be able to 
understand the magnitude of the vev without inserting parameters by
hand \cite{flatguts}.

Whether or not there is a GUT, one expects to find 
some flatons which are 
gauge singlets. To understand the possibilities here, note first
that each flaton will be a complex field.\footnote
{In a supersymmetric theory all scalar fields are complex since the 
corresponding left- or right-handed spin-half field 
has two degrees of freedom.}. If there is a single flaton field, and 
it 
is charged under a global
$U(1)$ symmetry, then its vev will spontaneously break that symmetry.
The prime candidate for a global $U(1)$ symmetry is the
Peccei-Quinn symmetry associated with the axion, and indeed 
a flaton model for that symmetry has been proposed
\cite{flataxion}. With more flaton fields one can have higher global
symmetries, as we mention at the end of Section V in connection with
topological defects. 

However there is no need for a flaton field to be 
charged under a continuous symmetry; on the contrary,
it is quite reasonable on the basis of current thinking to expect
flaton fields which are 
charged under at most discrete symmetries. For them, the vacuum 
manifold consists of discrete points instead of an entire circle. 
The possibilities that we have in mind are of two kinds.
The first are some or all of the moduli fields
\cite{problem,dilaton,Banks,Randall,Steinhardt,Dine},
which are ubiquitous in versions of supergravity motivated by the
superstring. For the present purpose, a modulus may be defined as a 
field whose potential is exactly flat in the limit of unbroken 
symmetry.
It is widely supposed that one or more of the moduli is a flaton,
with a vev of order the reduced Planck mass 
$\mpl=(8\pi G)^{-1/2}=2.4\times 10^{18}\,{\rm GeV}$.\footnote
{The origin of a field is defined to be a fixed point of 
its symmetries. With all relevant fields at the origin the potential
has zero slope, where `relevant' means those fields that are coupled 
to 
the one in whose direction the slope is being defined.
We are assuming that this is the 
situation for the flatons (or at least that any couplings to other 
fields displace the minimum by an amount small compared with the vev).
In the case of a modulus there is more than one 
fixed point with the separation between fixed points of order
$\mpl$. In this context the vev may be defined as the distance to the 
nearest fix point, and the statement that the vev is of order 
$\mpl$ simply means that it is not close to any particular fixed 
point.
}
(Lower values may be possible so that 
GUT Higgs fields may be moduli \cite{flatguts}, but for
simplicity, the term `moduli' will be used in what follows
to denote only the case where the vev is 
of order $\mpl$. It is also possible that moduli have masses much 
bigger 
than $1\TeV$, in which case they are not flatons.)
Now consider the opposite possibility, of
gauge-singlet flatons whose potential is not flat even in the limit
of supersymmetry \cite{therminf,ewan}.
 Extensions of the Standard Model
can easily contain such fields, and their potentials will
resemble Eq.~(\ref{eq:potential}) below, with couplings 
perhaps or order 1. As a result, their vev is typically 
much smaller than the GUT scale, perhaps 
of order $10^9$ to $10^{11}\GeV$.

\section{Flaton cosmology}

The cosmology of flatons is a rather rapidly moving research area at 
the 
moment, but we briefly summarize it here following 
\cite{therminf,klslatest,andreipersonal}. For simplicity we generally
pretend that there is only one flaton field $\Phi$, 
whose potential is invariant under a $U(1)$
symmetry so that it depends only on the magnitude $|\Phi|$. We 
also however point to the differences that can occur in the more 
realistic case, where there may be several flaton fields and 
a higher symmetry or else no continuous symmetry at all.

In the early universe, the effective potential
of a flaton field is modified by its interaction. If the interaction 
with light particles is 
of gravitational strength, then the effective potential is expected to
receive an additional contribution of order $\pm H^2 |\Phi|^2$
\cite{dinefisch,coughlan,Ross,fvi}. 
If it is somewhat stronger one might perhaps expect
a contribution $\pm \alpha^2 H^2 |\Phi|^2$ with $\alpha$
significantly bigger than 1. One expects gravitational strength 
interactions for moduli with a vev of order $\mpl$, 
whereas one expects interactions stronger
by a factor perhaps $\mpl/M$ for a flaton with smaller vev $M$.
These statements refer to field values of order the vev,
as opposed to values near the origin which we discuss in a moment,
and they are based on the idea that the coupling to a light particle
is suppressed by a factor $\propto 1/M$.

The weak contributions just described can be of either sign.
In contrast, if the interaction is strong enough to lead to 
thermalization then one has the finite temperature correction to the
potential, which at least near the origin gives  
typically a positive contribution 
$\sim T^2|\Phi |^2$ to the mass-squared.\footnote
{We will not consider a strong coupling of form 
$\delta V_I\sim -\lambda_\chi \Phi^\dagger\Phi\chi^2 $ with
$\lambda\chi\sim +1$,
which would result in a negative \cite{weinberg}
temperature-induced mass term
$\delta V_T\sim -(\lambda_\chi T^2/12)\Phi^\dagger\Phi$. (These 
type of terms do not destabilize the potential since 
$+\lambda_\Phi (\Phi^\dagger\Phi)^n $ eventually stabilizes it.)} 
For a flaton field $\Phi$, one
expects the thermal contribution to be present in the
regime $|\Phi|\lsim T$ because the effective mass of particles
coupling to the flaton field is of order $|\Phi|$.
Section III below is devoted
to a detailed study of the thermal contribution.

As a result of these modifications to the potential, a flaton in the 
early universe will either be at the origin or will 
have a value which is large but not particularly 
close to its true vev. We discuss these possibilities in turn.

\subsection{Flaton field initially at the origin}

Let us suppose that a positive effective mass-squared
holds the flaton field 
at the origin prior to full reheating after ordinary inflation.
Then the finite temperature mass-squared $\sim T^2$
will typically hold it there 
until the temperature 
falls to a value $T\sub{end}\sim m_0\sim 10^2$ to
$10^3\GeV$, where $-m_0^2$ is the true mass-squared of the flaton
field at the origin.
Only then will it start to 
oscillate about the true vev. On the other hand, the energy 
density with the flaton field at the origin is
\be
\rho=V_0+\frac{\pi^2}{30} g_*T^4
\label{rhot}
\ee
where $g_*$ is the effective number of species.\footnote
{This estimate does not apply to non-thermalized particles 
(in particular moduli) whose energy may come to dominate that 
of the thermalized particles before thermal inflation commences.
In this case we would have an additional contribution to the energy 
density
(the oscillating moduli fields typically behave as massive particles).
This would somewhat complicate our analysis \cite{therminf}  
but the qualitative conclusions remain unaffected, 
so we will not consider them.}
Because $T\sub{end}$ is so low, the first term dominates
before the flaton field rolls away, leading to an era of
`thermal' inflation. The era starts when
$T\sim V_0^{1/4}\sim (mM)^{1/2}$, and it ends when $T=T\sub{end}
\sim m$, so the number of 
$e$-folds of thermal inflation is of order
$(1/2)\ln (M/m)\sim 9+(1/2)\ln (M/10^{10}GeV)$. 
As a result, thermal inflation cannot 
replace ordinary inflation; rather, it takes place if at all
after ordinary 
inflation has ended, at the low energy scale 
$V_0^{1/4}\sim (mM)^{1/2}\sim 10^6\GeV(M/10^{10}\GeV)^{1/2}$.

After thermal inflation ends the flaton field
starts to oscillate about the minimum of the effective potential,
so we enter an era of 
matter domination by the flaton particles. 
(As we discuss in a moment the minimum might initially be shifted 
from 
the true vacuum, but this effect can be ignored because the movement
of the minimum will be slow.)
If each flaton particle decays at the single particle decay rate
$\Gamma$, this era ends at the time $\Gamma^{-1}$.
In typical models one estimates
$\Gamma^{-1}=100\gamma^{-1}M^2/m^3$ where the numerical factor 
$\gamma$ is 
at most of order 1. At least the bulk of the decay products
thermalize promptly, so setting $\Gamma^{-1}$ equal to the expansion time 
$H^{-1}$ where $H$ is the Hubble constant we obtain the temperature just after 
flaton decay to be
\be
T_{\rm D}\simeq g_*^{-\frac{1}{4}} \Gamma ^{\frac{1}{2}} 
m_{\rm Pl}^{\frac{1}{2}}
\simeq 3 \gamma^{\frac{1}{2}} \left( \frac{10^{11}\,{\rm GeV}}{M} 
\right)
\left( \frac{m}{300\,{\rm GeV}} \right)^{\frac{3}{2}} \mbox{GeV}
\label{TD}
\ee
where $g_*$ is the effective number of species at temperature
$T_{\rm D}$, set equal to 100 for the numerical estimate.

In fact, the assumption that the flatons corresponding to
the oscillation decay individually at the one-particle decay rate
is not correct, because of 
non-linear relaxation effects
of which the most studied example is parametric resonance
\cite{brand,Kofman,klslatest,Shtanov,Boy,yosh}
(see also \cite{ovrut,Ross}). As soon as the flaton field starts
to oscillate at the end of thermal inflation, non-linear
relaxation drains off a significant fraction of the oscillation,
or in other words destroys a significant fraction of the 
corresponding flaton particles. They are replaced by
marginally relativistic scalar particles of whatever species
have a significant interaction with the flaton field,
including the flaton field itself. 

If nothing happens to the produced scalar particles they will become
non-relativistic after a few Hubble times, and are expected to decay
at their one-particle decay rate.
If, on the other hand, they thermalize then they turn into
highly relativistic radiation.               
At the present time it is not clear whether parametric resonance can
really create particles which thermalize successfully.
However, it {\em is } clear that
the flaton component of the produced particles cannot thermalize
because here one knows that the interaction is too weak.
Furthermore, one expects that the energy density of the produced flatons
will be a significant fraction of the total energy density
\cite{andreipersonal}. Thus, even if the other produced particles
thermalize promptly one expects that a significant fraction of
non-thermalized energy will remain, and that a significant fraction
of {\em that} energy will be in flaton particles.

Any thermalized radiation produced by parametric resonance will redshift 
away,
so independently of the details one expects that a
few Hubble times after the end of thermal inflation
the energy density is dominated by non-relativistic
scalar particles, including the flatons and perhaps other species.
Each species will decay at the single-particle decay rate, so
we expect eventually to find only the longest-lived species, which
dominates the energy density until it decays.
For simplicity we shall assume in
what follows that this species is the original flaton.
Then, the upshot of this discussion is that despite non-linear effects,
the eventual reheat temperature
is still the temperature $T_{\rm D}$ calculated from
the single-particle decay rate, as given by Eq.~(\ref{TD}).

\subsection{Flaton field initially displaced from the origin}

Now suppose that the flaton field
has a large value in the early universe.
In contrast with the first case, the field will not now be
in thermal equilibrium because its interaction with light particles
will be very weak (because otherwise the large value would generate
a large mass for the would-be light particle). 
A very crude model for the evolution of the effective potential 
of a flaton field $\Phi$ in the early universe is 
\cite{therminf,andreipersonal,ewanpersonal}
\be
V(|\Phi|)= m^2(|\Phi|-M)^2 + \alpha H^2 (|\Phi|-\Phi_0)^2
\ee
where $M$ is the true vev.
The second term
represents the effect of 
interactions in the early universe, and for the present purpose
we can suppose that $\Phi_0\sim M$ is a constant, so that the only 
time 
dependence comes from $H$.

First suppose that $\alpha$ is of order 1, which is the expected value
at least for a flaton such as a modulus with gravitational 
strength interactions (corresponding to $M\sim \mpl$).
In that case \cite{therminf},
the minimum of $V$ is practically at $\Phi_0$ until the epoch
$H\sim m$, after which it moves quickly to the 
true vev $M$. The flaton will have settled down to the minimum of $V$
before the end of ordinary inflation, so the conclusion is that 
if $\alpha$ is of order 1 the flaton field
starts oscillating about the true vev at the epoch $H\sim m$, with
initial amplitude $|\Phi_0-M|\sim M$. 

As soon as the oscillations start, non-linear 
relaxation effects will convert a significant fraction of the energy
in the oscillating flaton field.
Provided that the oscillation continues to take place
about the true vev, then the qualitative picture will not be altered,
just as we discussed already in the case of an oscillation
starting out at the origin.
However, because the initial oscillation
energy can now be bigger than the potential energy $V_0$ at the origin,
there is now the possibility that non-linear relaxation effects restore the 
symmetry, so that the
oscillation takes place about the origin
\cite{klslatest}. 
In that case the flaton field oscillation might give way
to some $e$-folds of inflation, which might be thermal inflation,
before the symmetry breaks \cite{klslatest}. 
If it occurs, this will happen when the energy density is
equal to $V_0\sim (mM)^2$, which is much less than the 
energy density $\rho\sim (m\mpl)^2$ at the epoch when the oscillation 
starts.

Finally, consider the case $\alpha\gg 1$. The evolution of
the flaton field  is now
quite different \cite{andreipersonal}, because the
movement of the minimum of the effective potential
is always slow on the timescale
of oscillations around it. As a result
the flaton is at all times close to the 
minimum of its effective potential, and the cosmological production 
of 
flaton fields is strongly suppressed. (To be precise, analytic 
\cite{ewanpersonal} and numerical \cite{andreipersonal}
estimates show that the amplitude of the oscillation about the minimum
is reduced exponentially compared with the case $\alpha\sim 1$.)

\subsection{The case of more than one flaton field}

So far we have considered only a single flaton species.
If there are several, then some may be initially trapped at the origin
and some may be displaced from it. Those which are displaced
start to oscillate
first (when the energy density is $\rho\sim(m\mpl)^2$, corresponding to
a Hubble parameter $H\sim m$), while those trapped at 
the origin give rise to 
thermal inflation. If several flaton fields are trapped at the 
origin, the simplest possibility
is that they all roll away at more or less the same epoch,
which seems reasonable because the negative mass-squared
is of the same order for all flatons, and so is the 
finite-temperature correction. 
If that happens, we essentially recover the 
case of a single flaton. However, the rolling away of the first
flaton (say) might trap the remaining flatons at the origin
through its interactions. 
One possibility is that the particles produced by 
non-linear relaxation promptly thermalize, in which case one may 
enter a 
second era of thermal inflation \cite{therminf}.
Alternatively the remaining flatons may be trapped by non-linear
relaxation \cite{klslatest}, which could also lead to a 
second epoch of inflation. 

In this paper, we are assuming for simplicity that 
only genuine flatons can roll away from 
the origin. It could happen that a field with
a flat potential, and zero true vev, couples to a flaton in such a 
way 
that its effective potential acquires a nonzero  minimum by virtue
of the zero value of this flaton. Such a field could temporarily 
act as a flaton, leading to a more complicated cosmology
\cite{ewan} which we shall not consider.

\subsection{The entropy crisis and its solutions}

If a flaton comes to dominate the energy density of the universe,
then the low `reheat' temperature Eq.~(\ref{TD}) makes it 
cosmologically fatal if its vev is too large. 
Indeed, successful nucleosynthesis requires 
$T\sub{D}\gsim 10\MeV$, corresponding to 
$M \lsim 10^{14}\,{\rm GeV}$,
and thermalization of a stable LSP requires 
$T\sub{D}\gsim 1\GeV$ corresponding to 
$M\lsim 10^{12}\,{\rm GeV}$. 
(To have electroweak baryogenesis 
requires $T\sub{D}\gsim 100\GeV$ corresponding to
$M \lsim 10^{10}\,{\rm GeV}$, but this requirement
is not mandatory especially as thermal inflation itself
provides additional possibilities for baryogenesis
\cite{klslatest,ewan}.)

This possible problem is a concern for any moduli which are flatons
with a vev $M\sim\mpl$, and it is also a concern for GUT Higgs fields
if they are flatons since the corresponding vev is $M\sim
10^{16}\GeV$. It was originally termed the 
`entropy crisis' \cite{problem}, and was widely discussed in the
1980's \cite{dinefisch,coughlan,yam,therm,Enqvist,Ross,yam2,interm}.
More recently, focusing on moduli, it has been termed the
moduli problem \cite{dilaton,Banks,Randall,Steinhardt,Dine}.

Let us consider the status of the `moduli' problem, in the light of 
the 
above snapshot of the current status of flaton cosmology.
First, it should be emphasised that the problem exists only if the
fields in question are actually flatons, with
masses $\sim 10^2$ to $10^3\GeV$. Assuming that this is the case,
the status of the problem depends on whether or not the flaton in question 
is initially displaced from the origin. If it is, the
problem may not occur at all, because the flaton 
may settle down smoothly to the true vacuum without appreciable
oscillations; this may well occur for the GUT Higgs though
it is less likely for the moduli \cite{andreipersonal,ewanpersonal}.
If it does occur it can be solved by a subsequent
bout of thermal inflation, associated
with a flaton having 
a cosmologically safe vev $M\lsim 10^{12}\GeV$. 
As we remarked earlier, typical extensions of the 
Standard Model indeed contain such flatons.

 If, on the other hand, the flaton in question is initially
at the origin, then thermal inflation will occur and when it ends
the flaton will certainly oscillate with large amplitude.
In that case one needs a second era of thermal inflation 
\cite{therminf}, 
again associated with some other flaton which has a cosmologically 
safe vev. At least naively this in turn requires prompt 
thermalization of a 
significant fraction of the original flaton oscillation energy (with 
non-linear relaxation presumably first converting the energy into
marginally relativistic particles) and at the present
time it is not clear whether this is possible.
If it is not possible, then we reach the important conclusion that 
flatons with dangerously large vevs, like moduli and GUT Higgs fields,
cannot be associated with thermal inflation. 
We return to this issue for GUT Higgs fields in Section V.

\section{The 1-loop thermal effective potential}

In this section and the following one, 
we examine carefully the finite
temperature effects that 
are supposed to make thermal inflation possible.
They were first described by Yamamoto in 1986 \cite{yam}, 
but his account was
brief and it did not discuss the validity of the various 
approximations
employed. To our knowledge no subsequent authors have improved on 
Yamamoto's discussion.

As already remarked, the potential of the relevant flaton fields 
might be
invariant under a continuous symmetry such as a global $U(1)$ or a
GUT gauge symmetry, or alternatively it might have no continuous 
symmetry.
The computations in this paper will be applied to a simple toy model,
in which there is a single flaton field with a $U(1)$ symmetry.
Following \cite{gutti,therminf}, we take the potential 
to be of the form
\be
V = V_0 - m_0^{2}\Phi ^\dagger \Phi + \frac{\lambda_{n}}
{\mpl^{2n}} \left (\Phi ^\dagger\Phi \right ) ^{n+2}
\label{eq:potential}
\ee
where $n$ is an integer power. 
Here $m_0 $ is of order $10^3\GeV$,  and for 
simplicity we have included only the leading non-renormalizable 
term. The magnitude of the
coupling $\lambda_n$ is model dependent, but for definiteness 
one can keep in mind a value of order $1$.
We can parametrize the quantum field $\Phi$ as follows:
\be
\Phi = \frac{\phi}{\sqrt 2}+\delta \phi + i\eta, 
\ee
where $\phi$ is the classical degree of freedom, which equals to the 
expectation value of $\sqrt 2 \langle \Phi\rangle$, 
and $\delta \phi$ and $\eta$ are quantum fluctuating
fields with {\it zero\/} expectation values. 
The classical part of the effective
potential can be then written as
\be
V = V_0 - \frac12 m_0^{2} \phi ^2 + \frac{\lambda_{n}}
{\mpl^{2n}} \left (\frac{\phi ^2}{2}\right ) ^{n+2}
\label{eq:potentialII}
\ee
From this expression we can easily obtain  the mass $m_\phi$ 
of the flaton particle given by 
$m^2_\phi = d^2V/d\phi^2(\phi=\sqrt 2 M)$, 
the vacuum expectation value $M=\langle \Phi\rangle _0$, corresponding
to the minimum of $V$,
and the height $V_0$ of the potential at the origin which one gets
by setting $V(M)=0$. The result is
\bea
m^2 _\phi &=&2(n+1)m_0 ^2\\
\label{eq: m phi}
M^{2 (n+1)} &=& 
\frac{1}{\lambda_n}\frac{1}{n+2}
\; \mpl^{2n}\;\; m_0^2
\label{vev}\\
V_0&=& \frac{n+1}{n+2}\;M^2\,m_0^2
\eea

One naively expects $\lambda_n\sim 1$ ($n\ge 1)$ 
for the couplings, making
$M\sim 10^{10}\GeV$. On the other hand, it might be that the Planck 
scale is replaced by the GUT scale if there is one, which effectively
increases $\lambda_n$ leading \cite{therminf} 
to $M\sim 10^9$. Finally, $\lambda_n$ might be a small mass ratio,
which could make $M\gg 10^{10}$ so that it becomes \cite{therminf}
the GUT scale (appropriate if $\phi$ is a GUT Higgs field) or the Planck 
scale (appropriate if $\phi$ is a modulus). In this paper we focus 
on the first and second cases.

\subsection{Coupling to bosons}

Now consider the finite-temperature correction to the effective
potential. First we will argue that the thermal correction
from the $\Phi$ field itself is irrelevant in the sense that it can
neither trap the field at the origin nor cause a phase transition. 
Then, in order to cure 
this we will assume either that $\Phi$ couples to another 
(real) scalar field {\it via\/} a quartic coupling of form 
$g\Phi^\dagger\Phi \chi^2$, or gauge the U(1) by adding the
corresponding gauge field. We shall also consider the effect of 
coupling to
a spin-half field since we have in mind the case of supersymmetry.

The one-loop thermal correction 
of a bosonic excitation with the mass $m$ to the
effective potential has the following generic form
\cite{dolan}
\be \label{gfpot}
V_1 (m, T)= \frac{T^4}{2 \pi^2}
\int_{0}^{\infty} dx x^2  \ln \left( 1 - e^{-\sqrt{x^2 + 
m^2/T^2}}\right)
\label{eq: finite temperature}
\ee
Below we will always quote a one loop result for $m$. One should
keep in mind that strictly speaking in order to obtain 
the one loop thermal correction to the effective potential from
Eq.~(\ref{eq: finite temperature}) one should use the tree level
value for $m$. In some cases the ring-improved one loop
potential (which is a result of resumming certain so-called ring 
diagrams to all orders) is more accurate. It corresponds to using 
the one-loop thermal 
value for the effective mass {\it only\/} for the 
`singular' term $-m^3 T/12\pi$ in the high temperature
expansion. 

The one-loop value for the masses follow:
$m_\phi^2\simeq -m_0^2+gT^2/12$ is the mass-squared of
the $\Phi$-field excitations, and 
$m_\chi^2=m_\chi ^2 (T=0)+gT^2/12+g\phi^2$ is the mass-squared of 
$\chi$.
Note that in this approximation $m_\phi$ does not acquire a
$\phi$-dependent correction. Indeed, the lowest order correction
to $m_\phi^2$ proportional to $\phi^2$ occurs at the $(n+1)$-loop 
level and 
is of order $\lambda_n T^2 (T/\mpl)^{2n}$, which is minute and 
hence 
negligible. We now see from Eq.~(\ref{eq: finite temperature})
that the finite temperature contribution due to the $\Phi$ field 
excitations is to an excellent approximation a 
$\Phi$-independent shift in the effective
potential. This shift is nothing but adding a constant to the energy
density and hence it should not affect the dynamics of the $\Phi$ 
field.

The complete effective potential is now 
$V(\phi, T )=V(\phi)+\sum_i V_1(m_i,T)$, where we are summing over all 
the boson fields. 
In the regime $m_i\simeq \sqrt g \phi>> T$, the finite temperature
correction $V_1$ is negligible. (As we remarked earlier, one does not
in any case expect thermalization in this regime, since particles
with mass $m\ll \phi$ must couple weakly to 
$\phi$
\cite{therminf}.) In the opposite regime
$m_i\simeq \sqrt g \phi\ll T$, the correction is 
$V_1\simeq - T^4\zeta_4/\pi^2$ with $\zeta_4=\pi^4/90$, 
which is independent of $\phi$. The conclusion is that at
large temperatures $T\gg M$, the 
potential has a single minimum at $\phi=0$. 
(Recall that $M$ is the position of the true vacuum, defined by
Eq.~(\ref{eq: m phi}).) As the temperature drops
a second minimum develops. At a critical temperature $T_c$ 
defined by  $V(T_c,\phi=0)=V(T_c,\phi=\sqrt 2 M)$ the two minima are 
degenerate.\footnote
{Notice that this is not the temperature $T_{\rm C}$ defined in
\cite{gutti,therminf}; in this paper we are denoting that quantity
by $T\sub{end}$.}
This temperature is given by
\bea
T_c&=&\left( \frac{n+1}{n+2}\frac{\pi^2}{\zeta_4} \right)^{1/4}
(M m_0)^{1/2} \sim V_0^{1/4}
\label{eq: critical temperature} 
\eea

The bump in the effective potential which causes the phase 
transition is solely due to the $\chi$ field excitations.
Even though there are two degenerate minima the field will be trapped
at $\phi=0$, {\it i.e.\/} there will be no phase 
coexistence since the bump is very large and the tunneling rate is 
minute, much smaller than the expansion rate of the Universe. A
large bump
at the critical temperature indicates that the phase transition is 
strongly first order. As the Universe cools down below $T_c$, trapped 
in the false vacuum phase $\phi=0$, the false vacuum energy starts 
dominating. Using Eq.~(\ref{rhot}), this occurs at the temperature  
$T_{TI}^4 \simeq V_0 \pi^2/3g_*\zeta_4$,  at which 
Thermal Inflation begins. The 
effective number of particle species
$g_*$ is  
equal 3 in our toy model,
two from the $\phi$ field and one from the $\chi$-field.
In general $g_*$ will be of order $10^2 - 10 ^3$. 
Note that, according to the rough estimate Eq.~(\ref{rhot}), the 
temperature $T_c$ is of the same order of magnitude as the 
temperature $T_{TI}$ at which thermal inflation begins.

Thermal inflation ends when the Universe supercools sufficiently 
so that the tunneling rate for creation  of the true 
vacuum bubbles becomes comparable to the expansion rate of the 
Universe. We will postpone a more detailed study of 
the bubble nucleation to the next section.

So far we have focussed on the coupling to a spin zero 
particle $\chi$. If the flaton is a GUT Higgs field
there will also be a coupling to gauge particles. As a toy
model, consider the Abelian Higgs model, whose Lagrangian density
is ${\cal L}= -F^{\mu\nu}F _{\mu\nu}/4 
+(D_\mu\Phi)^\dagger (D^\mu\Phi)-V$, where
$D_\mu=\partial_{\mu} + ie A_\mu$, $A_\mu$ is the vector field, and 
$F^{\mu\nu}$ is the field strength. The
thermal correction to the potential has again the form of
Eq.~(\ref{eq: finite temperature}) but with the masses
squared $m^2_\phi=-m_0^2 +(3e^2) T^2/12$ for both the physical
excitation and the unphysical Goldstone boson of the $\phi$-field,
and $m_L^2=e^2 \phi^2 + (e^2/3) T^2$, $m_T^2=e^2 \phi^2 $ for (one)
longitudinal and (two) transverse polarized gauge field excitations, 
respectively. At this level some gauge dependence may come into the
play,  but we will not discuss this issue here.
The above analysis caries through in almost exactly the same way,
except
that the masses in the thermal potential are now
$m_L$, $m_T$ and $m_\phi$. There are 5 degrees of freedom relevant
for the
onset of thermal inflation, so the temperature at which it begins
will
be somewhat lower, as will be the critical temperature.

What about the validity of the one-loop approximation?
The squared coupling of the three dimensional theory is $g_3^2=gT$
for a scalar theory and $g_3^2=e^2T$ for a gauged theory.
Each loop added costs an extra factor $gT/M$; recall that 
$M_\chi\simeq \sqrt g \phi$, $M_{T,L}\simeq e\phi$, so that we expect 
the one-loop approximation to break down when $g_3^2> M$, which means
$\phi/T< \sqrt g$ for the scalar theory and  $\phi/T<e$ for the
gauge theory. In other words, the one-loop approximation can be 
trusted 
as one approaches the critical temperature up to 
$|T-T_c|\sim gT,\; e^2T$. In this case,
unlike in the case of the electroweak theory, the one-loop treatment
is accurate. The reason is that the 
maximum of the effective potential (bump) is located at 
\begin{equation}
\phi_{1} \sim \frac{T}{\sqrt g}
\label{eq: bump}
\end{equation}
This should be taken only as an estimate since 
the bump is located where neither the high temperature nor the low 
temperature expansion is accurate. Nevertheless, this estimate 
tells us something important about the validity of the one loop 
approximation: the bump is located in the region of validity of 
the one-loop approximation
(this is, for example, not true in the electroweak theory);
it is $1/g$ times larger for the scalar theory ($1/e^2$ for the gauge 
theory) than the value of $\phi$ at which we expect 
the breakdown of the one-loop approximation. 
We see that, at the local maximum, the 1-loop potential is still a 
reasonable approximation to the true effective potential
and this treatment of the phase transition is 
quite accurate. One can also evaluate the potential energy 
density at the bump.  To get a feeling for what it is we 
evaluate it in the high temperature limit and find 
$V(\phi_1) \sim V_0-T^4 \pi^2(1/90 -1/648)$. Hence we see that 
the energy density of the bump $(T^4\pi^2/648)$ is much smaller than 
the energy density difference between the false and true vacua 
$(V_0-T^4\pi^2/90)$. This suggests that the wall of the critical 
bubble is thick, which will be used in the following section where we 
study nucleation. 

\subsection{Coupling to fermions}

To illustrate what happens in the case of a scalar field coupling 
with fermions, we include in the global U(1) potential a Yukawa 
coupling of the scalar to a 
spinor field  $y_{f} \phi \bar{\psi} \psi$.
We will then have a contribution to the effective potential of the 
form
\be
V_{1}^f(m_f,T) =
-\frac{T^4}{2 \pi^2} \int_{0}^{\infty} dx x^2 \ln \left( 1 + 
e^{-\sqrt{x^2 + 
m^2_{f}/T^2}}
\right)
\ee
where the fermion mass is $m_{f}(\phi) = y_{f} \phi$. This yields
a critical temperature very close to the scalar coupling case of 
Eq.~(\ref{eq: critical temperature}) with a minor numerical 
difference: since each fermion contributes to the free energy $7/8$
times less than a boson, with the same number of degrees of freedom,
$T_c$ will be higher by a factor $(8/7)^{1/4}$. The local 
maximum $\phi_{1}$ will also be the same (replacing $e$ by $y_{f}$).
One should note an important difference between this phase transition 
and 
the electroweak phase transition. Since the main reason for the 
free energy valley at around $\phi\sim 0$ 
is not the singular cubic term $-m_f^3T/12\pi$ in
the high temperature expansion of Eq.~(\ref{eq: finite temperature}), but 
the exponential suppression of the population density for the mass
$m_f>>T$, the only difference in studying the cases of coupling to 
bosons and fermions is the trivial one of counting 
degrees of freedom. The total effective number of degrees of 
freedom is
$g_*({\rm tot})=\frac{7}{8}g_*({\rm fermion})+
g_*({\rm boson})$.
This remark will be also valid when studying the nucleation problem
in the next section. 

\section{Tunneling}

We are taking the standard route as advocated by Callan, Coleman 
\cite{callancoleman}, and Linde \cite{Linde}
in which we assume that the spherical three dimensional action
\begin{equation}
S_3=\int d^3 r \left [ {1\over 2} (\nabla\phi)^2+V(\phi,T)   
\right ]
\label{eq: potential eight}
\end{equation}
determines the nucleation rate per unit time and volume
\be
\frac{\Gamma}{V}=
m^4 \left (\frac{S_3}{\pi T}\right )^{3/2}{\rm e}^{-S_3/T}\,,\qquad
m^2=m_\chi^2\simeq \frac{gT^2}{12} + g\phi^2
\label{eq: nucleation rate}
\ee
This is an approximate formula but it is sufficiently accurate for 
our purposes. A more accurate treatment of bubble nucleation can be
find in \cite{CarringtonKapusta}. 
The rate in Eq.~(\ref{eq: nucleation rate}) has to be compared with 
the expansion rate of the Universe to determine the rate of 
bubble formation. More precisely \cite{MooreProkopec},
assuming a constant bubble expansion velocity, at any given time 
the fraction of the Universe remaining in the symmetric phase 
is given by
\be
\exp\left ( -\int _{-\infty}^t \frac{4\pi}{3} v^3 (t-t')^3 
\frac{\Gamma}{V} \, dt' \right )
\label{eq: fraction of symmetric}
\ee
In order to solve for the nucleation temperature we expand $S_3$ 
about the 
value when the integrand is about {\it one:} 
$S_3 = S_3(T_{nucl})+ (t-t')(dS_3/dt)$, where $d/dt=(dT/dt)d/dT$, 
$dT/dt=-TH$, and $H^2=(8\pi G/3) (\pi^2 g_*(T) T^4/30)$ is the Hubble 
constant.
Using a saddle point approximation we find that the integral is 
of order one when 
\be
{\rm e}^{S_3 (T_{nucl})/T_{nucl}}=\frac{8\pi v^3}{ (HdS_3/dT)^4}
 m^4\left (\frac{S_3 (T_{nucl})}{\pi T_{nucl}}\right )^{3/2}
\label{eq: nucleation temperature}
\ee
In order to solve this equation we have to find what is the bubble 
action $S_3$ as a function of temperature.

The solution to the (spherically symmetric, classical) equation of 
motion for $\phi$:
\begin{equation}
{d^2\over dr^2}\phi +{2\over r}{d\over dr}\phi=
{d\over d\phi}V (\phi, T)
\label{eq:potentialnine}
\end{equation}
with the boundary conditions:
\be
\frac{d\phi}{dr} | _{r=0} =0\,,\qquad \phi (r\rightarrow \infty)=0
\label{eq: boundary conditions}
\ee
specifies the wall profile. The bubble action is 
\be
S_3= 4\pi\int r^2 \, dr \left [ \left (\frac{d\phi}{dr}\right )^2
+  V(\phi, T)- V(0, T)\right ]
\ee
Since Eq.~(\ref{eq:potentialnine})
is analytically intractable, various approximation schemes have been 
developed. In general when the height of the local maximum is small 
(large)
in comparison to the false vacuum energy, the thick (thin) bubble 
wall 
approximation gives a reasonably accurate answer to $S_3$. 
First one can show using the thin wall approximation, which is 
accurate
around $T_c$, that the three dimensional action:
$S_3\sim (2\pi/3) \sigma^3/V(0,T)^2$  is indeed huge; here 
$\sigma=\int_0^{\phi;V(\phi, T)-V(0,T)=0} 
\sqrt{2 [ V(\phi,T)-V(0,T)]}d\phi\sim T_c^3/\sqrt g$ is the surface
tension of the bubble, and near the critical temperature 
$V(0,T)=\frac{\zeta_4}{\pi^2}(T_c^4-T^4)<<T_c^4$.

When the temperature drops significantly below $T_c$, such that 
$V_{\rm bump}<< V(0,T)$, where $V_{\rm bump}$ is the difference between the 
maximum of the potential and $V(0,T)$, 
the thick wall approximation becomes appropriate.
Here we present a heuristic discussion. The three dimensional 
action for the critical bubble can be approximated as follows:
\begin{eqnarray}
S_3 & \sim & 2\pi R_c \phi ^2 -
\frac{4\pi}{3} R_c^3 \left [ V(0, T)- V(\phi, T) \right ]\\
 V(0, T)- V(\phi, T) & \simeq & \frac 1 2 \frac{m_0^2 
T^2}{g}\varphi^2 
+ 
T^4 \left (\frac{\varphi}{2\pi}\right )^{3/2} \, {\rm e}^{-\varphi}
- T^4 \frac{\zeta_4}{\pi^2}\,,\qquad \varphi=\frac{\sqrt g \phi}{T}
\end{eqnarray}
where $\phi$ is the field value at the centre of the bubble. 
The critical action is extremal with respect to $R$ and $\phi$:
\be
\frac{\partial S_3}{\partial R_c}=0\,,\qquad
\frac{\partial S_3}{\partial \phi _c}=0
\ee
The first condition gives for the size of the critical bubble and the 
action:
\be
R_c^2\simeq \frac{\phi^2}{2 \left [ V(0,T)-V(\phi, T) \right 
]}\,,\qquad
S_3\simeq \frac{4\pi}{3}
\frac{\phi^3}{\sqrt{2 \left [ V(0,T)-V(\phi, T) \right ]}}
\ee
which can be re-written as 
\be
S_3\sim \frac{4\pi}{3g}\frac{T^2}{m_0}
\frac{\varphi^3}{\sqrt{\varphi^2-2(\sqrt g T/m_0)^2 (\zeta_4/\pi^2)
+2 (\sqrt g T/m_0)^2 (\varphi/2\pi)^{3/2} \exp -\varphi }}
\label{eq: three d action}
\ee
The variation of $S_3$ with respect to $\varphi$ gives then the value 
of $\varphi$ at the origin of the bubble:
\be
\varphi^2\simeq \frac{3\zeta_4}{\pi^2}
\left (\frac{\sqrt g T}{m_0}\right )^2 
\ee
which is obtained by neglecting the exponentially small term in 
Eq.~(\ref{eq: three d action}) so that the critical action is 
\be
s_3=\frac{S_3}{T}\simeq \frac{4\sqrt 3\zeta_4}{\pi}
\left (\frac{T}{m_0}\right )^3
\label{eq: critical action}
\ee

We can now use this result and 
Eq.~(\ref{eq: fraction of symmetric}) -- 
(\ref{eq: nucleation temperature}) to obtain the 
nucleation temperature $T_{nucl}$: 
\footnote{Note that $T_{nucl}$ for all practical purposes 
coincides with the temperature at the end of thermal 
inflation $T_{end}$, discussed in section 2 above.}
\be
s_{nucl}\equiv \frac{S_3 (T_{nucl})}{T_{nucl}} \simeq
-6+4\ln\frac{m_{Pl}}{M}+\frac{1}{6}\ln \frac{S_3}{T}
\simeq 17+4\ln \frac{10^{16}GeV}{M}
\label{eq: bubble action}
\ee
where we assumed that the bubble is supersonic:
$v\sim 1$; this is plausible since supercooling is very large. 
temperature is about $T_{nucl}\simeq 2 m_0$. 
For somewhat different values of parameters $T_{nucl}$ 
changes only slightly.
To a very good approximation $T_{nucl}$ is proportional to $m_0$. 
(Here we have neglected a logarithmic dependence of the form 
$(\ln m_0 )^{1/3}$.)

These results are in substantial agreement with
that  of Yamamoto \cite{yam}, who found that the 
bubble action is (up to $\ln m_0$) proportional to $T^4/m_0 ^3$. 
What we have done is to look more 
carefully at the somewhat delicate assumptions which have to be made 
in order to 
derive it. We have found that at a reasonable level of confidence they
can indeed be justified by carefully considering the regime of 
validity of the various approximations.

\section{Topological defects with a flat potential}

Now we come to the second part of our investigation, which 
is to look at topological defects associated with a flat potential.
In the case of a single (complex) flaton field with a $U(1)$ symmetry,
which we have focussed on for the sake of simplicity, these are 
cosmic strings. If instead there is only a  discrete $Z_n$
symmetry they are domain walls. Finally, if there are two or more
flaton fields there could be a higher continuous symmetry, which,
depending on the pattern of symmetry breaking, could lead to 
monopoles and/or textures (or to no defects at all).
We shall consider all these cases.

In order for the defects to form, the symmetry has to be restored at 
early times, which as discussed in Section 2 may or may not happen
for a given flaton field. Assuming that it is indeed restored
the defects form when the field rolls away from the origin, and since we 
are considering a flaton field this occurs at the end of an era of 
thermal inflation. Thus, we are considering the formation of topological 
defects at the end of such an era.

We shall focus particularly on
the case where there is a GUT whose Higgs fields 
have a flat potential, 
schematically of the form Eq.~(\ref{eq:potential}).
As mentioned already such a flat potential seems quite natural,
because
one can then hope to generate the required vev from scales already 
present in the theory \cite{flatguts}. For example, in 
Eq.~(\ref{eq:potential}), 
 the coupling $\lambda_n$ might be of the form
$(m_{\tilde g}/\mpl)^p$, where $m_{\tilde g}$ is related to vevs 
arising from gaugino condensation, and $p$ is a positive integer.
This paradigm is of course very different from 
the usual one, where the 
potential of the GUT Higgs is supposed to be non-flat so that it has 
the 
form
\be
V=\lambda (\Phi^\dagger\Phi-M^2)^2
\label{nonflat}
\ee
with $\lambda\sim 1$.
Such a form is indeed natural for the Standard Model Higgs where
$M$ is only of order the susy breaking scale $10^2$ to $10^3\GeV$,
but it seems far less natural for a GUT. In other words, there 
seems to be no reason to suppose that the vev of a 
GUT Higgs is of the same order of magnitude as its mass.

GUT symmetry breaking with a flat potential is completely different
from  the usually considered case. 
The GUT phase-transition
occurs at the end of an era of thermal inflation. Just before it
occurs, the potential energy density $V_0$ of the Higgs field accounts
for most of the energy density in the universe, and after the transition
this energy is initially converted into a homogeneous oscillation of the
Higgs field (away from any topological defects), corresponding to
extremely non-relativistic Higgs particles. Through non-linear 
relaxation
effects this initial era is quickly followed by one in which
at least a significant fraction
of the energy density resides in
marginally relativistic scalar particles, of various species including
the Higgs particles. There may also be a significant fraction in
thermalized (and hence highly relativistic) particles, but this
question has not been settled at the present time.
                      
Because the potential is flat, the GUT Higgs particles are light
(mass $m\sim 10^2$ to $10^3\GeV$), and according to Eq.~(\ref{TD})
they decay after nucleosynthesis. In order not to upset nucleosynthesis
their abundance must be diluted by 
a second era of thermal inflation.\footnote
{Or perhaps
a non-thermal analogue \cite{klslatest}, but we shall not consider that
possibility.}
Such an era requires the thermalization of a substantial fraction of the
Higgs particles, and, taking the fraction to be of order 1, but not very 
close to it, the abundance of the Higgs particles immediately
after thermalization is 
$n/s\sim V_0^{1/4}/(g_*^{1/4} m)\sim 10^6$. From \cite{constraint}, their
abundance at the epoch of nucleosynthesis must satisfy $n/s\lsim 10^{-12}$,
so the second bout of thermal inflation should 
dilute the abundance by a factor of at least $10^{18}$.

The actual dilution factor actually provided by thermal inflation is
\cite{therminf}
\begin{equation}
\Delta \sim  \frac{\epsilon}{75} \frac{m^2 M^2}{T\sub{D} T\sub{end}^3}
\sim 10^{18}\left( \frac{M}{10^{11}\GeV}
\right)^2 \left( \frac{1\GeV}{T\sub{D}} \right)
\label{eq: second bout of inflation}
\end{equation}
In this expression $m$ and $M$ refer to the second era of thermal
inflation, as do $T\sub{end}$ (the temperature at 
the end of thermal
inflation) and $T_D$ (the reheat temperature).
The quantity $\epsilon\leq 1$ is the fraction of energy left
unthermalized
after non-linear relaxation effects have occurred after the end of the
second bout of thermal inflation, which we set equal to 1.

In making the above estimates we have set $V_0=m^2M^2$ and 
$T\sub{end}=m=10^{2.5}\GeV$ for both eras of thermal inflation.
The expected range for the vev of the inflaton causing the
second bout of thermal inflation 
is $10^9\GeV\lsim M\lsim 10^{12}\GeV$ (with the lower limit coming from
the discussion after Eq.~({\ref{vev}) and the upper limit from the 
requirement of decay before nucleosynthesis).
Since several parameters have been set equal to 
fiducial values, a second bout of thermal inflation seems well able
to provide the necessary dilution. The only problem is to ensure that
it occurs, bearing in mind the fact mentioned earlier that 
the possibility of thermalization immediately after thermal inflation
ends has not been demonstrated.

Assuming that the GUT particles can be somehow diluted, 
we proceed to a discussion of the defects produced, starting with
monopoles.
                   
\subsection{Monopoles}

It is well known that monopoles form when the vacuum manifold $M$ of a 
gauge group $G$ contains non-shrinkable 2 surfaces, {\it i. e.\/}
when $\pi_2({\rm M})\neq I$, where $\pi_2$ is the second homotopy group. 
In the light of a theorem in homotopy theory, this may be rephrased as:
monopoles form whenever a grand unified 
semisimple group $G$ breaks down to a group H which has 
nontrivial fundamental homotopy group $\pi_1({\rm H})\neq I$,  {\it 
i.e.\/} H contains non-contractable loops. A simple example 
is an H which contains at least one U(1) as a factor, just like the 
Standard Model! 
It is argued in \cite{JeannerotDavis} that since supersymmetry does 
not change the connected structure of the (super)-Lie group, 
the same consideration as above is valid for a supersymmetric GUT.
Since rather generically the symmetry of a low energy 
theory contains non-contractable loops, one must check what would be
the cosmological implications of monopoles' formation, {\it i.e.} 
whether there is a monopole problem.

It is convenient to specify the monopole abundance $n_{mon}$ as a fraction
of the entropy density $s$, because $n_{mon}/s$ is constant as long as
monopoles are not destroyed, except when particle decay (or some other
non-equilibrium process) increases the entropy per comoving volume.
Consider first the observational upper bound on the present monopole
abundance. 
The mass of a gauge monopole for a given vev $M$ is not much affected 
by the 
flatness of the potential, because it is known that the mass
goes to a finite limit as the potential becomes flatter and flatter
(Bogomolny bound) \cite{vs}. This leads to a 
simple and reliable bound on the present monopole abundance 
of $n_{mon}/s\lsim 10^{-24}$, coming from the requirement that 
the monopole density be less than critical.
 This bound is stronger than
that derived from the requirement that monopoles do not affect
nucleosynthesis. The next bound one can derive is based on the
requirement that the galactic magnetic field not be dissipated
through acceleration of monopoles from the time of galaxy
formation \cite{Parker}. Strictly speaking this bound applies 
only to the monopole abundance in galaxies. The 
original Parker 
bound is $n_{mon}/s\lsim 10^{-24}$. This was argued to be too
strong; when the depletion of monopoles, as they get accelerated
and ejected from the galaxy, is taken into account 
\cite{LazaridesShafiWalsh}, the bound becomes 
$n_{mon}/s\lsim 10^{-20}$. A stronger bound was found using 
the fact that a grand unified monopole catalyzes nucleon decay
\cite{CallanRubakovWilczek}, when captured by a neutron star;
the bound is $n_{mon}/s\lsim 10^{-30}$ \cite{KolbColgateHarvey}.
A more stringent bound can be derived based on monopole
capture and consequent catalytic nucleon decays
in the main sequence stars \cite{vs}: $n_{mon}/s\lsim 10^{-36}$.
The flatness of the potential is not expected to
alter this bound because the strong interaction cross section for
the monopole is independent of its radius 
\cite{CallanRubakovWilczek,daviscat}.

In order to find out whether there is a monopole problem, we 
ought to compare the above bounds with the amount of 
monopoles that form at the phase transition. 
Since we need a second epoch of thermal inflation 
(to dilute the GUT Higgs particles), a significant fraction of the 
energy density had better thermalize immediately after the GUT Higgs
phase transition, leading to a temperature $T\sub{GUT}\sim
(mM)^{1/2}\sim 10^9\GeV$. (Here $m\sim 10^2$ to $10^3\GeV$ is the
mass of a typical GUT Higgs particle with a flat potential, and $M
\sim 10^{16}\GeV$ is the vev of a typical GUT Higgs field.)
This is much bigger than the temperature
$T\sim m$ just before thermalization, but it is 
still far below the temperature $T\sim M$ which one would
obtain after thermalization with a non-flat potential.
In the case of a non-flat potential the temperature is high enough to 
initially annihilate the monopoles, so that when annihilation ceases 
there is of order one monopole per horizon volume irrespective of the 
initial abundance. For  a flat potential this is not the case, and 
we have to consider the initial abundance.

A lower limit on it is provided by 
the causality bound which states that on average about
$p\simeq 1/8$ monopoles form per causality volume 
$V_c\simeq (4\pi/3)H^{-3}$, where $H^{-1}$ is the causality 
radius. The probability $p\simeq 1/8$ can be obtained from an argument
based on the Kibble mechanism: the field is uncorrelated on 
superhorizon scales, and the probability that the randomly 
oriented field  in different correlation volumes will cover
more than half of the vacuum manifold is about $p=1/8$.
This bound gives an initial monopole density 
$n_{mon}\simeq (3/4\pi) p H^3$, corresponding to 
\begin{equation}
\left (\frac{n_{mon}}{s} \right )_{caus}  
\gsim \frac{8p}{g_*^{1/4}}
\frac{(m_0 M)^{3/2}}{m_{Pl}^3}\sim 10^{-27}
\label{eq: one monopole per horizon}
\end{equation}
We have used for the entropy density
$s\simeq g_*^{1/4} \epsilon_{conv}^{3/4}(m_0 M)^{3/2}$;
here  $M$ denotes the GUT Higgs vev, $m_0$ the curvature of the 
potential at the origin and $\epsilon_{conv}$ is the fraction of the energy in 
the field that decays and thermalizes. The numerical 
estimate in 
(\ref{eq: one monopole per horizon}) is obtained by 
setting $m_0\sim 10^3$GeV, and
$M\sim 10^{16}$GeV. 

To obtain an actual estimate, remember that we
have argued that the Universe undergoes a strong supercooling and 
the phase transition is strongly first order proceeding via 
bubble
nucleation. An estimate  
of the typical bubble size at collisions will allow us to estimate
monopole production. Since each bubble nucleates with a 
random phase at the vacuum manifold, when four bubbles
collide a monopole (or an anti-monopole) forms with the
probability $p\sim 1/8$. The only missing information is now 
the typical bubble size when they collide. A typical bubble  
nucleates at the temperature $T_{nucl}\simeq 2m_0$, as specified by
(\ref{eq: critical action}) and Eq.~(\ref{eq: bubble action}).
The typical bubble size
can be obtained from Eq.~(\ref{eq: fraction of symmetric}) as 
follows.
The integral in the exponent is dominated by bubbles of size
\begin{equation}
l=v\tau
\sim 3\frac{1}{HdS_3/dT}
\simeq\frac{3}{4}\frac{T}{S_3}
\frac{1}{H}\simeq \frac{1}{20}\frac{1}{H}
\label{eq: bubble size}
\end{equation}
Hence, the typical bubble, when it collides, is about 1/20th of
the horizon size $1/H$. This means that we have an estimate for 
monopole production which is about $20^3\sim 10^4$ times above
the causality bound in Eq.~(\ref{eq: one monopole per horizon}):
\begin{equation}
\left (\frac{n_{mon}}{s} \right )_{bubble}  
\sim 10^{-23}
\label{eq: best estimate of monopole formation}
\end{equation}
According to this estimate the dilution factor provided by the second 
bout of thermal 
inflation is ample.

In the discussion so far we have supposed that the monopoles are 
isolated,
but in many cases they  are connected by cosmic strings
\cite{copeland,vs}. 
If this is the case then strong effective forces will be induced by
connecting monopoles and anti-monopoles with strings which will cause 
efficient annihilation all the way to the horizon scales, 
establishing 
a scaling solution. (In this respect the cosmology of the strings 
connecting monopoles resembles the cosmology of global monopoles 
discussed below. This does not mean that monopoles connected with strings 
may affect structure formation, as it is in the case of global monopoles
discussed below, 
since the ordering field dynamics of global monopoles is very
different than that of the magnetic monopoles. In these models,
if any cosmologically interesting effects exist, they would be
produced by the strings.)  
On scales larger than the horizon causality 
prevents annihilation. Nevertheless, this annihilation mechanism will
suffice to keep the monopole density at the level of a few per horizon
at all times and hence solve the monopole problem \cite{copeland,vs}. 

On the assumption that there is no monopole problem, let us proceed
to a discussion of cosmic strings.

\subsection{Cosmic strings}

Cosmic strings form in the case when the vacuum manifold $M$ of the 
broken theory contains non-shrinkable loops so that 
$\pi_1($M$)\neq I$. In other words (using some homotopy theory),
they form if a semisimple grand unified 
group $G$ breaks down to a group H which is not connected, 
{\it i.e.\/}  $\pi_0({\rm H})\neq I$. (Recall that 
connectedness of a group is not changed by supersymmetry.)
Since the symmetry groups of low energy theories are often connected, cosmic
strings are a less generic feature than
monopoles, but they do occur for some choices of the GUT gauge group
\cite{JeannerotDavis}. (As an example, SO(10) may break 
down to the Standard Model enhanced by an additional matter parity
$Z_2$, which is, of course, {\it not}  connected.)

On scales below the horizon, the cosmic string network has a `scaling'
configuration which is more or less independent of its previous 
history.
As a result, the cosmology of the strings will not be significantly 
affected by the ten or so $e$-folds of inflation taking 
place after their formation. Instead, it will depend mainly on the 
string mass per unit length $\mu$. As is well known, 
the strings are candidates for the origin of both the
cmb anisotropy and large scale structure, with some
value $\mu\sim (10^{16}\GeV)^2$.
If this turns out to be the case then 
$\mu$ will be determined, and in any event one already has an upper
bound on $\mu$. In contrast with the more commonly considered model
of a smooth, Gaussian primordial perturbation, the cosmic string
prediction has yet to be accurately calculated over a wide range of 
scales, but as a guide we may take the prediction for the 
low multipoles of the cmb anisotropy.
Here it is known \cite{markcmb,hk,CoulsonFGT} 
$\mu\simeq 2\times (10^{16}\GeV)^2$, with an uncertainty of 
perhaps a factor 2 either way.\footnote
{One of the uncertainties in this estimate is the fact
that small-scale structure in the string network
breaks the equality of the
mass-per-unit length and the tension, by
a few tens of percent  \cite{vv91}.
The estimates
being referred to, on the other hand, are performed under the assumption
that these quantities have a common value $\mu$.}
If the effect is observed and is due solely to strings, one will 
therefore deduce a value of
$\mu$ somewhere in the range 
\be
\mu \simeq
1 {\rm \ to \ }4\times (10^{16}\GeV)^2
\label{result}
\ee
which imposes an upper limit
\be
\mu\lsim 4\times (10^{16}\GeV)^2
\label{bound}
\ee

What do these results for $\mu$ tell us about the scale of GUT
symmetry breaking, as defined by the vev $M$?
With a non-flat potential of the form Eq.~(\ref{nonflat})
it is known \cite{hk} that
$\mu\simeq \pi M^2$. Using this result, 
an observed signal in the cmb anisotropy
will imply a value 
\be
5\times 10^{15}\GeV \lsim
M\lsim 1\times 10^{16}\GeV
\label{mresult}
\ee
and one already has an upper limit
\be
M\lsim 1\times 10^{16}\GeV
\label{mbound}
\ee

We want to know how much this estimate is changed by 
the flatness of the potential, and in contrast with the case of the 
monopole mass the question is non-trivial because $\mu$ does not 
tend to a finite limit as the potential becomes flatter and flatter.

We work with the same toy model
that we considered earlier, namely the case of a U(1) gauge symmetry
and a potential of the form 
(\ref{eq:potentialII}) with $n=2$ ($\phi^8$ coupling).
The Lagrangian density of the flaton coupled to the U(1) gauge 
field is
\be
{\cal L} = -\frac{1}{4} F_{\mu\nu} F^{\mu\nu}  + |D_{\mu}\Phi|^{2} 
- V(\Phi)
\ee
where $D_{\mu} = \pd_{\mu} +ieA_{\mu}$, $F_{\mu\nu} = 
\pd_{\mu}A_{\nu} - 
\pd_{\nu} A_{\mu}$ and $V$ is the tree-level, zero temperature scalar 
potential.
The equations of motion in the Lorentz gauge ($\partial^\mu A_\mu=0$)
for this Lagrangian are
\bea
& & D^{2}\Phi + \frac{\pd V}{\pd \Phi^{*}} = 0 \\
& & \pd_{\nu} F^{\mu\nu} + ie(\Phi^{*} D^{\mu} \Phi -
(D^{\mu}\Phi)^{*} \Phi) = 0
\eea
We now make a (static) cylindrical ansatz for $\Phi$ and $A^\mu$ 
(setting, for 
simplicity, the winding number to one).
\be
\Phi = M \:  f (m_{V} \rho) e^{i\theta}
\;\;\;{\rm and}\;\;\;
\vec{A} = \frac{1}{e \rho} a (m_{V} \rho) \hat{\theta}
\ee
where $(\rho,\theta,z)$ are the cylindrical coordinates, $f$ and $a$ 
generic 
functions, $M$ is the vev of the $\Phi$ field (see equation 
\ref{vev}), and $m_{V} = e M$ is the 
vector particle mass. The equations of 
motion will then read: 
\bea
& & f'' + \frac{1}{\xi} f' -
\frac{1}{\xi^{2}} f (a-1)^2 -\beta f (f^{6} -1) = 0 
\label{eq: eom local strings}
\\
& & a'' - \frac{a'}{\xi}
+ 2 f^{2} (1 - a) = 0
\eea
where we have defined $\xi = \m_{V}\rho$ and 
$\beta = m_0^{2}/m_{V}^{2}$.

The asymptotic behaviour of the solutions to these equations are 
similar to the ones for the $\phi^4$ potential  
$$
f \sim \left\{ \begin{array}{l} f_{0} \xi, \\
1 - f_{1} \xi^{-1/2} exp(-\sqrt{\beta} \xi), \end{array} \right.
\;\;\;\;
 a \sim \left\{ \begin{array}{ll} a_{0} \xi^{2}, & {\rm as} \;\xi
\rightarrow 0; \\ 
1 - a_{1} \xi^{1/2} exp(-\xi), & {\rm as} \;\xi \rightarrow \infty. 
\end{array} 
\right.
$$

We now are interested estimating the energy per unit length, $\mu= 
\int 
\epsilon dV$, of these strings. The integral is over all space, and 
$\epsilon$ 
is 
the energy density, which in this case reads as
\be
\epsilon(\rho) = \left|
\frac{\pd \Phi}{\pd \rho} \right|^{2} + \left| \frac{1}{\rho}
\frac{\pd \Phi}{\pd \theta} - ieA_{\theta} \Phi \right|^{2} + V(\Phi)
+ \frac{1}{2} {\vec B}^{2} 
\ee
where $\vec{B} = \vec{\nabla} \wedge \vec{A}$ is the
magnetic field. With our ansatz, the energy per unit length becomes
\be
\mu= \pi M^{2} \int \xi d \xi \left( f'^{2} + \frac{1}{\xi^{2}} 
f^{2}
(1-a)^{2} + \frac{1}{6} \beta \left[ \frac{3}{4} + f^{2} (\frac{1}{4} 
f^{6}-1)
\right] +\frac{a'^{2}}{\xi^{2}} \right)
\label{eq: local string energy}
\ee
In the usual case, with $\beta \sim 1$ and the quartic coupling, the 
integral is 
close to one (it can actually be shown that it is one when $\beta = 
1$) and the 
mass scale of the string is given by the vacuum vev 
\cite{bogomolnyi76,jacobs79}. One expects a 
similar 
result 
here, but our coupling is much smaller than usual with
$\beta \sim 10^{-27}$. It
is not very difficult to estimate the energy per unit length and the 
profile of 
these static vortices numerically. Figure 1 shows the results 
thus 
obtained for a 
large range of values of the $\beta$ parameter for both the 
quartic and 
$\phi^8$ cases. The energy decreases logarithmically with small 
$\beta$ and so 
the very small coupling changes the value of $\mu$ by only a small 
factor of two 
orders of magnitude. The difference between the two types of 
potential is also 
seen to be small (around $20 \%$ for $\beta \sim 1$) and getting 
smaller with 
decreasing $\beta$.

that when $\beta$ is very small, $\mu$ becomes rather insensitive
to the value of $\beta$, with a value $\mu\sim (0.09 -0.14)\times 
M^2$ (to be compared with the value $\mu\simeq \pi M^2$ for 
a non-flat potential with $\beta\sim 1$).
This estimate applies in a rather large range of the coupling 
$\beta\sim 10^{-20}-10^{-35}$. It follows that an observed signal in 
the
cmb, corresponding to the range defined by Eq.~(\ref{result}), 
will correspond to
\be
3\times 10^{16}\GeV \lsim M\lsim  6.5\times 
10^{16}\GeV
\ee
which converts to an upper bound on $M$ of
\be
M < 6.5\times 10^{16}\GeV
\ee

For a given GUT theory, the scale $M\sub{GUT}$
at which the three Standard Model couplings unify is determined
by their values measured at low energies.
In simple GUT models, $M\sub{GUT}$ is found to be around $2\times
10^{16}\GeV$, and with a single Higgs field  one would have 
$g M\simeq M\sub{GUT}$ where $g$ is the gauge coupling at the 
unification point. This coupling in turn is typically given by
$g^2/(4\pi)\sim 1/25$, corresponding to $g\sim 1$, so that 
with a single Higgs field one would have $M\simeq M\sub{GUT}$. 
However, in a realistic model there will be more than one GUT Higgs 
field and all of the analysis in this subsection will require 
modification. 

Taking into account the present uncertainties,
the GUT unification scale required to make 
strings a candidate for the origin of large scale structure
is consistent with the one deduced from low-energy
data, both for a flat and for a non-flat potential.
In particular, the upper limit on the unification scale
deduced from the cmb anisotropy be no bigger than
that observed by COBE is consistent for both forms of the 
potential. On the other hand, it is clear that 
in the future one or other form of the potential might be preferred, 
since we have seen (Fig.~1) that 
the ratio $\mu/M^2$ is significantly different for the two cases.

\begin{figure}
{\hfil 
\epsfysize=6cm
\epsfbox{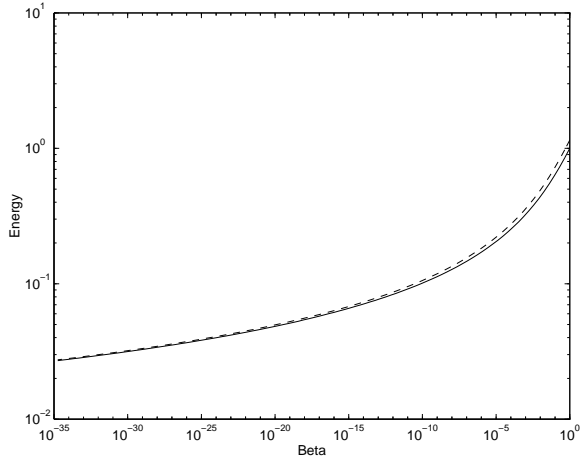} 
\hfil}
\caption{$\mu$ as a function of $\beta$}
\small{
The energy per unit length of the static vortex solution for the 
$\phi^4$ 
potential (full line) and the $\phi^8$ potential (dashed line).
The energy per unit length 
is in units of $\pi M^2$.} 
\end{figure}

\subsection{Global symmetry defects}

Although a GUT is initially constructed to satisfy a gauge symmetry,
global symmetries could occur accidentally and
be spontaneously broken by fields with vevs at the same
scale $\sim 10^{16}\GeV$. Global symmetries, notably the
Peccei-Quinn $U(1)$ symmetry have also been proposed at
lower energies. In addition to the possibilities of 
monopoles and strings, there are now the possibilities
of domain walls and textures. We discuss textures first.

\subsubsection{Textures}

It is known that textures with a vev $M\sim 10^{16}\GeV$
can significantly affect the cmb anisotropy and large
scale structure \cite{turok}.
Again, the central question we want to answer is whether
the flatness of the potential alters this situation.
As with strings, the configuration approaches a scaling solution
so the only thing we have to worry about is the energy of the
textures for a given vev $M$. The scaling solution for textures 
is characterized by continuous texture formation on the 
horizon scales
and 
their subsequent collapse, as governed by the scalar field
ordering dynamics. 

It turns out that the relevant mass scale is not affected by the 
flatness
of the potential. We will now argue that this is indeed so.
The size of the self-coupling $\lambda$ does not matter 
as long as it is large enough to confine $\Phi$ to the vacuum given 
by 
$|\Phi|=M$. Since we are interested in the epoch relevant for 
structure 
formation, the relevant scale is $T_{dec}\sim 1 eV$, which is much 
below any scale in the
microscopic theory. Indeed the `flatness scale' is given by 
the curvature of the potential at its minimum
$m=V''(M)^{1/2}\sim m_0\sim 1\TeV>> T_{dec}$.
A more rigorous argument can be constructed by looking at the 
evolution equation for textures in an expanding Universe. This 
equation has an effective self-coupling 
re-scaled as $\lambda_{eff}\sim (a(T)/a_0)^2\lambda$
\cite{TurokSpergel}, where 
$a$ is the scale factor of the Universe.
At $T\sim T_{dec}$, the ratio $a(T_{dec})/a_0\simeq
T_{GUT}/T_{dec}$ is so big that $\lambda_{eff}\sim
(m_0/T_{dec})^2>>1$ is still very large. This means that 
the effective coupling for all practical purposes is large and 
the field is confined to $|\Phi|=M$. This implies that the 
non-linear $\sigma$-model  \cite{TurokSpergel}
is a very accurate description of the texture dynamics, even for flat 
potentials. Normalising to the $10^{o}$ scales reported by COBE, and using 
the cosmological parameters $\Omega =1,\,\Lambda =0$ and $h=0.5$ the result 
may be read off from \cite{CoulsonFGT}:
$M\simeq 1.4\times 10^{16}$GeV. 

A possible realization of a texture model within grand unified 
theories
is proposed in \cite{JoyceTurok}
in which the GUT  of the form
$G\times SU(3)_{fam}$ (where $SU(3)_{fam}$ is 
an additional global family symmetry) has been considered. 
A supersymmetric GUT group with a flat Higgs potential 
may have a similar form. The family symmetry $SU(3)_{fam}$ 
breaks with the GUT, so it is plausible 
to assume that the texture potential is also flat. 

\subsubsection{Global monopoles and domain walls}

The dynamics of global field ordering 
with SO(3) symmetry (global 
monopoles) can be studied using similar numerical techniques as 
developed for textures \cite{CoulsonFGT}.
Also the ordering dynamics can be 
to a good approximation represented by a non-linear
$\sigma$-model \cite{BennettRhie}, \cite{CoulsonFGT}. 
They eventually reach a scaling solution
in which `monopoles' form on the scales of the horizon and then 
collapse
to the core size given by $\sim 1/m_0$. Since the interactions
between monopoles are strong and long range, there will be 
a strong tendency for monopole-anti-monopole annihilation 
which will keep the number of monopoles (anti-monopoles) at the level 
of order one per horizon at any time. In that respect, the monopole 
dynamics
resembles the dynamics of textures, with an important difference of 
an occasional monopole -- anti-monopole annihilation event, which 
will tend to imprint a strong non-Gaussian signal on the cmbr.
Recalling the assumptions made in their calculation concerning the values 
of the cosmological parameters we have read off from \cite{CoulsonFGT} that 
global monopoles
have an observable signature on large scale structure and/or the
cmb anisotropy  for some vev of order
$M\simeq 1 \times 10^{16}$GeV.

For walls we can make a simple estimate based on 
the energy per unit area of the wall: 
$\sigma\simeq \int dz [ (d\phi/dz)^2/2 + V(z)]$. Making 
the simple substitutions $d/dz\sim 1/l$, $\int dz\sim l$, $V(z)\sim 
V_0
\sim (m_0 M)^2$, we have $\sigma\sim M^2/l+V_0 l$, which is 
extremized when $l\sim M/\sqrt {V_0}\sim 1/m_0$ so that 
the energy per unit area is about $\sigma\sim M^3 (m_0/M)$,
where $m_0\sim 10^3\GeV$. Compared with 
the usual case one sees that walls are much fatter, but less massive.
However, the reduction in mass is far too small to allow them to 
be cosmologically viable. Instead they had better either not form
or disappear harmlessly, which is quite possible since the
discrete symmetry leading to their appearance will probably 
be slightly broken if it is not a gauge symmetry \cite{therminf}.

\section{Conclusion}

The topics discussed in this paper relate to a considerable body of
ongoing research at the moment, whose common theme is the form of the 
effective potential in the early universe, and the consequent
cosmological effects of the scalar fields. Some of this research
continues to address the traditional, and still very important,
question of how to implement an early era of ordinary (slow-roll) 
inflation.
Here we have focussed instead on aspects of `thermal inflation',
which occurs if at all after ordinary inflation and lasts
for only a few $e$-folds. 

The fields which can give rise to thermal inflation are 
`flaton' fields. By definition, these fields have
`flat' potentials and `large' vevs, where these terms
refer to the mass scale $10^2$ to $10^3\GeV$.
They arise naturally in currently favoured extensions of the 
Standard Model, along with the opposite case
of fields with `flat' potentials but zero vevs.

A flaton field gives rise to thermal inflation if it is trapped at 
the 
origin by virtue of the finite temperature correction to its 
effective 
potential, after the thermal contribution to the energy density
has become comparatively small. According to a calculation made
a decade ago by Yamamoto, this trapping is expected to occur for a 
flat
potential (provided that the flaton field is indeed in the vicinity 
of the 
origin), with thermal inflation ending only when the local minimum of 
the 
effective potential at the origin almost completely disappears.
Part of our objective was to investigate carefully the rather delicate
assumptions needed to arrive at this conclusion, using some modern
perspectives particularly on the tunneling rate out of the false 
vacuum.
Happily, we confirm the conclusion.

The actual cosmology of flaton fields, which determines whether a 
given flaton field  will actually find itself in the
vicinity of the origin so that thermal inflation can take place,
is at present a rather rapidly moving research area. 
We have not attempted any new advance here, contenting ourselves
with a brief snapshot of the situation as it stands at present.
Instead we have addressed a question which has received relatively
little attention, which is the nature and cosmology of topological 
defects forming at the end of thermal inflation. 
One would expect {\it a priori} that their 
cosmology might be significantly affected by the 
flatness of the potential. 
The answer to this question turns out to depend
on the type of defect.

A particular focus for the second part of our investigation has
been the possibility that there is a GUT, whose Higgs fields
have a flat potential. This possibility is quite natural, and it 
leads to a 
cosmology very different from the usually considered case of a 
non-flat
Higgs potential. GUT symmetry breaking, if it occurs, will 
be preceded by an era of thermal inflation. The Higgs particles
produced at the transition are cosmologically dangerous because they 
are light and long-lived, and to dilute them
one needs a second era of thermal inflation.
%But if one presumes that there was a way resolving the Higgs
%remnants problem without invoking a second bout of inflation, then 
%we find that monopoles are produced at the right level to be
%a dark matter candidate. This amount of monopoles is compatible 
%with the Parker bound, but violates the bounds based on catalytic 
%decays of nucleons in neutron and main sequence stars. If these 
%bounds are taken seriously, a second bout of thermal inflation 
%would resolve the conflict. 
This period of thermal inflation will also dilute the monopole
abundance sufficiently, but 
{\it will not eliminate
cosmic strings} because it lasts only a few $e$-folds.
A simple picture of a string network evolution in an inflationary
Universe is as follows. First as the Universe expands
the strings will quickly reach the density of about one (long) string
per horizon volume. After that the network freezes out since there
exists no causal process that could incite nontrivial dynamics. This
means that from then on the average correlation length grows
exponentially as the expansion rate, 
{\it i.e.\/} as $\xi(t)=\xi(t_0)\,{\rm e}^{H(t-t_0)}$. 
After about 10 $e$-foldings the correlation length has not grown more
than $\xi\sim {\rm e}^{10} /H$; this scale will come back within
the horizon after $\ln (a/a_0)\sim 10$ which in radiation era means 
at the temperature $T=T_D\times {\rm e}^{-10}
\simeq T_D\times 10^{-4.3}$; after
this strings will quickly reach the radiation era scaling solution. 
This temperature is far above the relevant scale for 
onset of cmbr anisotropies and structure formation,
keeping the strings cosmologically interesting. 

With all this in mind, 
we expect that cosmological relevance of cosmic strings is 
specified solely by the mass per unit length of strings.
We find that the GUT Higgs vev needed for the strings to give an 
observable signature in the cmb anisotropy is
 a few times bigger than in the case of a non-flat potential,
but still compatible with the 
typical estimate
$\simeq 2\times 10^{16}\GeV$ of the scale at which the 
Standard Model gauge couplings become unified.
Perhaps the main message is that
the flatness of the potential has surprisingly little effect on the
string cosmology.

We also discuss global defects, and in particular focus on the case 
of 
global textures and monopoles. We arrive to a rather surprising 
conclusion
that global texture and monopole dynamics is not affected at all by
the flatness of the potential. This has a simple explanation: the 
potential is simply not flat enough! The `flatness scale' 
$m_0\sim 10^3$GeV is large in comparison to the scale relevant for 
cosmology: $T\sim 1$eV, so the flat potential 
suffices to confine late textures to the vacuum manifold.
This leaves both textures and global monopoles as a viable candidate
for structure formation with the scales of symmetry breaking 
given by $M\sim 1.4\times 10^{16}$GeV and 
$M\sim 1\times 10^{16}$GeV, respectively. Finally in passing we 
give an argument that domain walls are still a problem. 

In summary, the study of fields with flat potentials and
large vevs is becoming another field on which particle physics 
increasingly
meets cosmology. Our hope is that either cosmology will put 
constraints 
on particle physics models, or particle physics will offer some 
interesting cosmological phenomena, at best tell us something about 
the origin of structures in the Universe.
Since this area of research is developing very
fast, it is very likely that in near future one will be able to make 
more definite statements about the formation and nature of defects in 
theories with flat potentials, and hence make a more definite 
prediction
on their cosmological relevance.

\subsection*{Acknowledgements}

We thank David Bailin, Beatriz de Carlos, Anne Davis, 
Andrei Linde, Graham Ross and Ewan 
Stewart for useful discussions.
T.P. acknowledges support from PPARC and NSF. T.B. acknowledges support from 
JNICT (Portugal).

\end{document}